# Shear SAW piezoelectricity of single crystalline $LiNbO_3$ films grown on 42°YX-$LiTaO_3$ substrates by DLI-CVD

Mikhail Mirushchenko,[1] Roland Salut,[1] Ludovic Gauthier-Manuel,[1] Vincent Astié,[2] Samuel Margueron,[1] Jean-Manuel Decams,[2] Yuancheng Ji,[3] Farshad F. Bolamiri,[3] Ventsislav Yantchev,[3,4] and Ausrine Bartasyte[1,5,6*]

[1] *Institut FEMTO-ST, Université Marie et Louis Pasteur, ENSMM-SUPMICROTECH, CNRS (UMR 6174), 26 rue de l'Epitaphe, 25030 Besançon, France*

[2] *Annealsys, 139 rue des Walkyries, 34000 Montpellier, France*

[3] *Terminal Antenna and RF Lab, Huawei Technologies Oy, 9 Itämerenkatu, 00180 Helsinki, Finland*

[4] *Technical University of Sofia, TMM/FIT, 8 Kliment Ohridski Blvd., 1756 Sofia, Bulgaria*

[5]*C2N, CNRS, Université de Paris-Saclay, 10 Boulevard Thomas Gobert, 91120 Palaiseau, France*

[6] *Institut Universitaire de France, 1 rue Descartes, 75231 Paris, France*

*Correspondence: ausrine.bartasyte@femto-st.fr

**Abstract**

Direct growth of single-crystalline $LiNbO_3$ films with technologically relevant nonpolar orientations and single ferroelectric domain state remains challenging. Here, we demonstrate the epitaxial growth of a 320-nm-thick, single-domain 42° YX-oriented $LiNbO_3$ film on a congruent 42° YX-cut $LiTaO_3$ substrate by direct liquid injection chemical vapor deposition. The deposition temperature was limited to 600 °C to preserve the ferroelectric domain state of the $LiTaO_3$ substrate. X-ray diffraction confirmed that the $LiNbO_3$ film inherited the substrate orientation and domain state and exhibited no detectable secondary phases. The LN rocking-curve full width at half maximum was 0.011°, approaching the instrumental resolution and demonstrating its high crystalline quality. Single-port leaky surface acoustic wave (LSAW) resonators with a wavelength of 4 µm were fabricated on the $LiNbO_3/LiTaO_3$ heterostructure and on a bare $LiTaO_3$ reference substrate. At approximately 1 GHz, the $LiNbO_3/LiTaO_3$ based resonators exhibited electromechanical coupling coefficient of about 13 %, compared to 8.5% for the reference $LiTaO_3$ based resonators. These results demonstrate that direct epitaxial growth can produce high-quality, strongly piezoelectric $LiNbO_3$ films and provide a route toward thickness-controlled layered substrates for high-performance acoustic devices.

Keywords: SH-SAW, films, CVD, $LiTaO_3$, $LiNbO_3$

Radio-frequency (RF) filters are essential components in information and communication technologies, including mobile phones, wireless data transmission, Wi-Fi, and digital subscriber lines, as well as in automotive navigation and tolling systems, medical instrumentation, industrial equipment, and household appliances.[1] Currently, $LiNbO_3$ (LN) and $LiTaO_3$ (LT) thin films are primarily produced from bulk single crystals using top-down techniques, such as mechanical polishing for film thicknesses above approximately 400 nm or ion slicing and wafer bonding for thicknesses between approximately 300 nm and 1 μm. However, the polishing steps involved in top-down fabrication make it difficult to achieve uniform and precisely controlled film thicknesses. The resulting thickness local variations can cause deviations in the device response, reduce reproducibility, complicate large-scale manufacturing, and require time-consuming frequency-trimming processes.

Compared to LT, LN provides stronger electromechanical coupling ($K^2$) but exhibits a larger negative temperature coefficient of frequency (TCF). Accordingly, submicrometer thick LN layers are employed in longitudinal leaky surface acoustic wave (LL-SAW) and Lamb wave resonator-based RF filters with enhanced bandwidths, [2, 3] while submicrometer thick LT layers are widely used in shear horizontal (SH)-SAW resonator-based RF filters characterized by very low loss and moderate $K^2$.[4] The direct growth of LN (or LT) films on LT (or LN) substrates or directly on top-down fabricated films, could enable $K^2$ and TCF values to be tailored between those of the two materials. Directly grown LN and LT layers could also provide improved thickness control and uniformity, potentially below 1%. This approach may be particularly useful for adjusting the thickness of ion-sliced or polished films, thereby overcoming some of the thickness limitations inherent to top-down fabrication.

For many years, the direct growth of high-quality LN and LT films was considered particularly challenging because of the difficulty in controlling the $Li_2O$ content and, consequently, the composition and physical properties of the films.[5] Nevertheless, the direct growth of high-quality, single-crystalline, c-axis-oriented LN films on sapphire substrates has been demonstrated, together with high-performance guided SAW propagation at 5.4 GHz.[6] However, c-axis-oriented LN and LT films do not provide the strongest $K^2$ for acoustic-wave devices; films with an in-plane or inclined c-axis are therefore of particular interest.[5] The growth of textured 33° YX-cut LN films (denoted (YXl)/33°),[7] on bottom Pt electrodes has been reported with physical properties approaching those of single-crystalline material.[8,9] This orientation enables the pure excitation of a pure longitudinal bulk acoustic wave (BAW) with the strongest $K^2$ predicted for LN. Although such textured films are compatible with BAW resonators, they are unsuitable for SAW and Lamb wave devices, which require single-crystalline films with well-defined in-plane orientation. [10,11] To date, satisfactory acoustic performance has not been demonstrated for directly grown single-crystalline LN or LT films with an inclined or in-plane c-axis.[5,12] The measured $K^2$ coefficient of SAW devices based on such films has remained below 1%, despite the substantially higher values predicted theoretically.[5,12] Even less is known about the direct growth of technologically important LN and LT orientations such as for example 41-42°YX-cut, which are widely used in SAW devices[10] and whose surface is not parallel to a low-index crystallographic plane. It is also noted that crystal orientations important for technology are those which provide strong piezoelectric coupling to a specific acoustic mode, while the other modes are significantly suppressed and separated from the main response. Only this type of resonator performance is suitable for filter and multiplexer design, where spurious mode content may seriously impact their performance. This indicates that further ameliorations of grown LN/LT film quality with technologically important orientations offering single-crystalline quality and single ferroelectric domain state need to be considered.

In this work, we demonstrate the epitaxial growth of a single-domain, single-crystalline LN film that inherits the crystallographic and ferroelectric-domain orientations of the underlying 42° YX-cut LT substrate. The resulting layered piezoelectric structure enhances the $K^2$ of leaky surface acoustic waves (LSAWs), which is known for its shear polarization and strong $K^2$ in both LN and LT. The properties of the grown LN film were assessed by comparing the performance of identically designed LSAW resonators fabricated on the layered LN/LT structure and on a bare LT reference substrate.

LN films were deposited by direct liquid injection chemical vapor deposition (DLI-CVD, MC100, Annealsys) on commercial congruent 42° YX-cut LT substrates at 7 Torr and nominal temperatures of 600 and 650 °C. Li(tmhd) and $Nb(tmhd)_4$ precursors (tmhd: tetramethylheptanedionate) were dissolved in mesitylene with orthophenanthroline as a Lewis base; further details are given in Refs. 13 and 14. Pyroelectric currents were measured on annealed and pristine LT substrates with sputtered Pt electrodes using an AIXACCT system and LINKAM heating stage, cycling around 52 °C at 4 °C/min. Film structure was characterized by X-ray diffraction (XRD) using a Bruker D8 Discover diffractometer with monochromatic Cu $K\alpha_1$ radiation ($\lambda$ = 1.54056 Å).

The LN surface was touch-polished to minimize roughness effects on SAW performance. Single-port resonators exciting X-propagating LSAWs (wavelength $\lambda$ = 4 μm) comprised a 30$\lambda$ aperture, 117 IDT (interdigital transducer) finger pairs, and 35 reflector electrodes on each side, with a designed pitch of 2 μm and mark-to-pitch ratio of 0.5. Ti/Al/Ti electrodes were fabricated by electron-beam lithography, evaporation, and lift-off. Focused-ion-beam-milled cross sections imaged by scanning electron microscopy (FIB–SEM) revealed thicknesses of 320 nm for LN, 14 nm for Ti, and 333 nm for Al (Fig. 1(a)), with an actual mark-to-pitch ratio of 0.49.

The epitaxial LN layer piezoelectric coupling was assessed using two-dimensional FEM simulations calibrated against measured resonator responses.[15] Starting from standard bulk constants for congruent LT and LN,[16] the model was fitted first to the bare LT reference and then to the LN/LT resonator by adjusting the material constants. Simulations incorporated the measured layer thicknesses, electrode widths, and sidewall angles. The resonators were characterized through admittance measurements over 0.5–1.5 GHz at room temperature using a FormFactor 250-μm-pitch Ground-Signal Z probe and a Keysight Streamline P9374A vector network analyzer after open-short-load calibration. The value of the effective device $K^2$ was estimated from the resonance frequency $f_r$, identified at the admittance maximum, and the antiresonance frequency $f_a$, identified at the admittance minimum:

$$K^2 = \frac{\pi^2}{4}\left(\frac{f_r - f_a}{f_r}\right) x100\%. \qquad (1)$$

Preserving the LT single-domain state requires a substrate temperature below Curie temperature of LT ($T_c \approx 603$ °C), while film crystallinity requires sufficient heating. Because the substrate is cooler than the thermocouple reading, nominal temperatures of 600 and 650 °C were tested without precursor injection. The pyroelectric current decreased from approximately 2 nA at 600 °C to 0.015 nA at 650 °C, indicating that the higher temperature exceeded $T_c$ and induced a polydomain state. A nominal deposition temperature of 600 °C was therefore selected to preserve the substrate single-domain state. The resulting LN film was dense and free of visible extended defects, such as grain boundaries, and exhibited a sharp interface with the substrate (Fig. 1(a)).

The epitaxial orientation of the LN films grown on 42° YX-cut LT substrates was determined by XRD. The surface of a 42° YX-cut LT substrate is not parallel to a low-index crystallographic plane. The closest low-index planes belong to the are $\{01\bar{1}2\}$ family and are inclined by 9.07° about the X-axis with respect to the substrate surface. Consequently, the 2Θ/ω scans and rocking-curve measurements were performed at a tilt angle of χ = 9.07°. Because the LN and LT $\{01\bar{1}2\}$ planes have very similar interplanar spacings, the film and substrate contributions were more clearly distinguished using their third-order $(03\bar{3}6)$ reflections in the 2Θ/ω scans [Fig. 1(b)]. The overlapping $(03\bar{3}6)$ reflections from the LN film and LT substrate were deconvoluted by fitting the experimental profile with Gaussian functions. The 2Θ offset caused by uncertainty in the sample height was then corrected by aligning the LT substrate reflection with the reference value of 76.285° reported for congruent LT by Santaro *et al.*[17] After this correction, the LN reflection was located at 76.093 ± 0.007°, in good agreement with the reference value of 76.088° reported for congruent LN by Abrahams *et al.*[18] The lattice parameters of the LN film may be influenced by both its Li stoichiometry and in-plane thermal and epitaxial stresses. Therefore, the Li stoichiometry cannot be determined reliably from the measured reflection position without a more detailed analysis of the residual stress and complementary compositional measurements.[19] Raman analysis was inconclusive because the stronger LT substrate signal masked the LN film response owing to overlapping vibrational modes. Although the precise Li stoichiometry could not be determined, the XRD measurements revealed no detectable secondary phases and confirmed that the LN films were epitaxial, single-crystalline, and followed the 42° YX orientation of the LT substrates. The rocking curve of the LN $(03\bar{3}6)$ reflection, measured by fixing 2Θ at the film-peak position identified in the 2Θ/ω scan, exhibited a full width at half maximum (FWHM) of 0.011° [Fig. 1(c)]. The FWHM of 0.013° obtained for the corresponding LT substrate peak in the same rocking-curve measurement does not represent the substrate mosaicity because 2Θ was fixed at the LN reflection position and was therefore offset from the LT Bragg condition. In contrast, the minimum LT rocking-curve width of 0.009°, measured with 2Θ fixed at the LT reflection position, represents the instrumental angular resolution in this angular range. Thus, the measured LN rocking-curve width of 0.011° approaches the resolution limit of the diffractometer and indicates extremely low mosaicity.

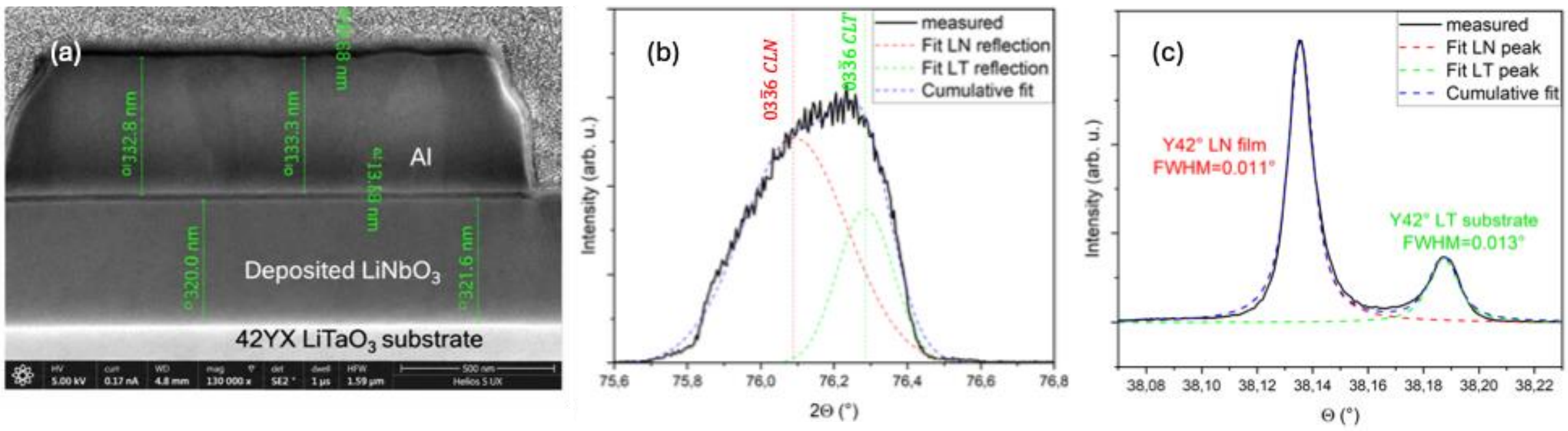

***Fig. 1.*** *(a) Cross-sectional SEM image of the FIB-milled SAW resonator stack, consisting of an LN film deposited on a 42° YX-cut LT substrate and a Ti/Al/Ti electrode structure. (b) ). 2Θ/ω scan (reference angles for congruent LN*[18] *and LT*[17] *are indicated by dotted lines) and (c) rocking curve of the* $(03\bar{3}6)$ *reflections of the LN/LT heterostructure, measured at a tilt angle of* $\chi$ *= 9.07° about the X-axis.*

The LSAW resonators on 42°YX-LN/LT heterostructure (propagation along X-axis) and on 42°Y-X-LT substrate were studied in a comparative manner. The acoustic response of the reference 42YX-LT substrate presented resonance frequency of 0.99 GHz and $K^2$ of 8.5%. The acoustic response of resonators on LN/LT heterostructure presented the resonance frequency of 0.98 GHz and $K^2$ of 12.9% for the studied LN layer thickness. Both measurements are shown together on Fig. 2. It is evident that the LSAW resonator on LN/LT heterostructure exhibit wider resonance to antiresonance frequency bandwidth, which highlights, for the first time, the strong piezoelectric coupling to SH-SAWs of the grown 42°YX LN thin-film. More accurate assessment of the LN shear wave piezoelectricity is done by means of comparative FEM analysis of the fabricated resonators. In Fig. 2 the calibrated FEM simulations are shown together with measurement data. Excellent fit between measurement and simulations is reached by slight tuning of the initial materials constants for the LT and LN, respectively. The LT reference substrate was subjected to the same thermal procedures as the substrate used in the LN/LT heterostructure, while no LN depositions were performed. For best fit to experimental data, the LT materials constant were shifted by few percent from the nominal. The measured device Q factors of the LSAW resonators were 1350 for the LT reference and 570 for the LN/LT heterolayer substrate. The observed Q degradation is well predicted by the 2D FEM analysis and is caused by energy leakage in the substrate bulk. The Q factor of the LT based resonator is in range of the industrial standards for this type of device, which in turn suggests good mechanical quality of the IDT electrodes fabricated with our technology. It is noted that the enhanced bulk wave radiation in the LN/LT heterostructure suggest different optimal crystal cut for achieving minimum losses. The latter however is not the scope of this work, which introduces a novel SAW resonator technological platform that requires further investigations. The FEM analysis provides further insight into the level of shear wave $K^2$ as compared to the LN bulk crystal. Within this fit the rotated LN elastic constant $c_{66}^{E,\mathrm{eff}}$ was reduced by 6% from nominal, the dielectric permittivity $\varepsilon_{11}$ was increased by 25% from nominal and the piezoelectric constant $e_{16}^{\text{eff}}$ was increased by 7% from nominal. Accordingly, the shear wave piezoelectric coupling of the LN layer was estimated $K_{SH,L}^2 = e_{16}^{\text{eff}2}/(c_{66}^{E,\text{eff}} * \varepsilon_{11}) = 0.97\ K_{SH,N}^2$ to be 97% from the nominal $K_{SH,N}^2$ of the congruent bulk crystal. It is important to note that the properties of LN are strongly affected by lithium nonstoichiometry. An increase in dielectric permittivity with increasing lithium content has been reported, corresponding to approximately 6% per mol% $Li_2O$.[20] For the (YX*l*)/42° orientation, the effective elastic stiffness is a linear combination of the crystal elastic constants:

$$c_{66}^{E,\text{eff}} = 0.55c_{66}^{E} + 0.45c_{44}^{E} + c_{14}^{E}\ . \qquad (2)$$

Using the elastic constants reported in Ref. 21 for congruent and nearly stoichiometric LN crystals containing 48.45 and 49.5 mol% $Li_2O$, respectively, $c_{66}^{E,\text{eff}}$ was estimated to be $75.95 \pm 0.37$ and $75.04 \pm 0.37$ GPa, corresponding to a decrease of 1.14% per mol% $Li_2O$. The dependence of the piezoelectric coefficient $e_{16}^{\text{eff}}$ on lithium content remains less established, although the available literature suggests that the piezoelectric coefficients of LN are sensitive to composition.[20] Attributing the changes in the estimated film constants solely to lithium content would imply, an apparent $Li_2O$ content above the stoichiometric value of 50 mol%. The reported $Li_2O$ concentrations in LN crystals may carry substantial uncertainties because they are estimated indirectly and depend on the stoichiometric reference used. Nevertheless, the observed trends in the dielectric and elastic constants suggest that the grown films are closer to the stoichiometric composition than congruent LN. Moreover, the properties of the grown films are affected not only by the Li nonstoichiometry but also by biaxial strain imposed by the LT substrate.

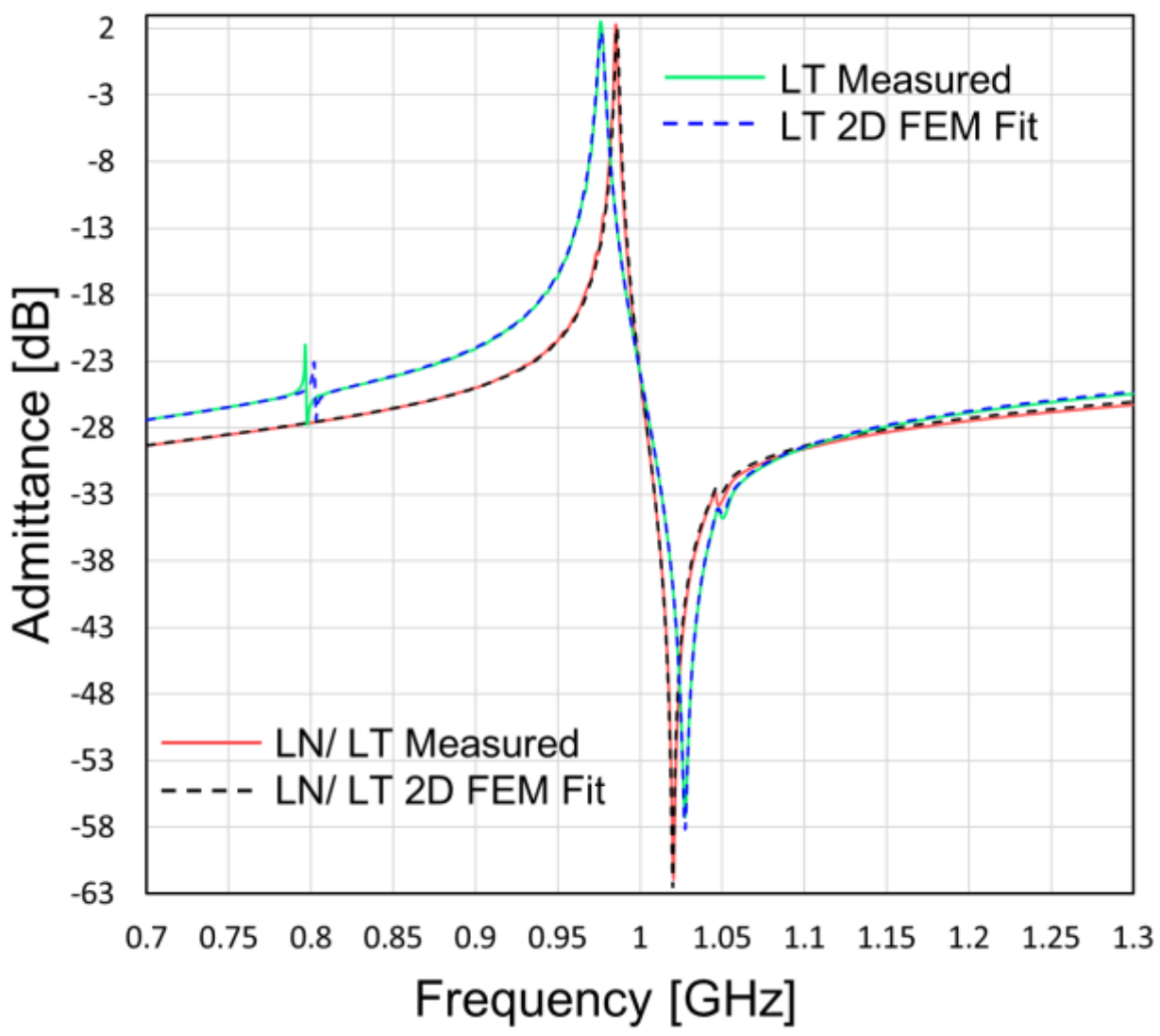

***Fig. 2.*** *Close to resonance admittance characteristics of identical design LSAW resonators on 42Y-X LT substrate and 42Y-X LN/LT heterostructure.*

In conclusion, we demonstrate the direct epitaxial growth of a 320-nm-thick, single-ferrolectric-domain LN film on a 42° YX-cut LT substrate using DLI-CVD. By controlling the deposition temperature to preserve the substrate ferroelectric domain state, we obtain a film that inherits its crystallographic and domain orientations. The film exhibits no detectable secondary phases and a rocking-curve width of only 0.011°, approaching the instrumental resolution. The central advance is the experimental demonstration of strong coupling to shear-polarized leaky surface acoustic waves (97% of $K^2$ of congruent bulk LN) in directly grown 42° YX-LN. We propose a new approach based on the epitaxial growth of single-domain LN films with a predetermined crystal orientation defined by the cut of the underlying LT substrate. The ability to form LN/LT layered piezoelectric substrates opens new avenues for research in microacoustic resonator technology. This approach is expected to enable the identification of promising LT substrate orientations for designing high-performance Rayleigh SAW and SH-SAW resonators based on LN/LT heterolayer piezoelectric substrates. Furthermore, dedicated layer transfer technology needs to be developed to fabricate a variety of piezoelectric-on-insulator substrates with inherently thickness uniform LN layers.

**Acknowledgments**

This research was funded by French RENATECH network, Bourgogne Franche-Comté region, EUR EIPHI program grant number ANR-17-EURE-0002, French National Research Agency funding ANR ANTARES (grant number ANR-24-CE51-0541) and ANR LINKS (ANR-20-CE08-0025), and the European Union's Horizon Europe research and innovation programme under the Marie Skłodowska-Curie Actions grant agreement No. 101169557 (HINA). V Yantchev acknowledges the support of BG16RFPR002-1.014-0006 project at TU-Sofia.

## References

[1] 5G Evolution and 6G, 2020 NTT DOCOMO, INC. Available: https://www.docomo.ne.jp/english/corporate/technology/whitepaper_6g/

[2] P. Turner, B. Garcia, V. Yantchev, G. Dyer, S. Yandrapalli, L. Villanueva, R. Hammond, V. Plessky, Electron. Letters 55, 942-944 (2019).

[3] P. Liu P, S. Fu, B. Xiao, X. Zhou, Q. Xu, J. Gao, S. Zhang, R. Wang, C. Song, F. Zeng, W. Wang, F. Pan, Microsyst Nanoeng. 11(1), 148 (2025).

[4] T. Takai, H. Iwamoto, Y. Takamine, T. Fuyutsume, T. Nakao, M. Hiramoto, T. Toi, and M. Koshino, Proc. IEEE Ultrason. Symp., 1-8 (2017).

[5] A. Bartasyte, S.Margueron, T. Baron, S. Oliveri, and P. Boulet, Adv. Mater. Interfaces 4 (8), 1600998 (2017).

[6] A. Almirall, S. Oliveri, W. Daniau, S. Margueron, T. Baron, P. Boulet, S. Ballandras, S. Chamaly, and A. Bartasyte, Appl. Phys. Lett.114, 162905 (2019).

[7] A. H. Meitzler, D. Berlincourt, G. A. Coquin, F. S. Welsh III, H. F. Tiersten, and A. W. Warner, IEEE Standard on Piezoelectricity, ANSI/IEEE Std Series 176-1978, Institute of Electrical and Electronics Engineers, New York, USA (1978).

[8] A. Bartasyte, V. Astié, and S. Kuprenaite Layered solid element comprising a ferroelectric layer and method for manufacturing the same, patent, EP20 305575 (2020).

[9] Q. Micard, V. Astié, L. La Spina, S. Oliveri, P. Boulet, S. Margueron, J.M. Decams, and A. Bartasyte, Proc. IEEE UFFC-JS, 1-3 (2024).

[10] A. Hagelauer, R. Ruby, S. Inoue, V. Plessky, K.-Y. Hashimoto, R. Nakagawa, J. Verdu, P. De Paco, A. Mortazawi, G. Piazza, Z. Schaffer, E. T.-T. Yen, T. Forster, and A. Tag, IEEE J. Microwaves 3 (1), 484 (2023).

[11] V. Yantchev, B. Garcia and V. Plessky, Transversely excited film bulk acoustic resonator using rotated Y-X cut lithium niobate, United States Patent 10790802 (2020).

[12] Y. Shibata, N. Kuze, K. Kaya, M. Matsui, M. Kanai, and T. Kawai, Proc. IEEE Ultrason. Symp., 247 (1996).

[13] A. Bartasyte, V. Plausinaitiene, A. Abrutis, S. Stanionyte, S. Margueron, P. Boulet, T. Kobata, Y. Uesu, and J. Gleize, J. Phys.: Condens. Matter 25, 205901 (2013).

[14] A. Bartasyte, S. Oliveri, S. Boujnah, S. Margueron, R. Bachelet, G. Saint-Girons, D. Albertini, B. Gautier, P. Boulet, I. Nuta, E. Blanquet, V. Astié, and J.M. Decams, Nanotechnology 35 (17), 175601 (2024).

[15] V. Yantchev, Y. Ji, T. Daiping, Y. Geng, J. Micromech. Microeng. 36, 073101 (2026).

[16] G. Kovacs, M. Anhorn, H.E. Engan, G. Visintini, and C.C.W. Ruppel, Proc. IEEE Int. Ultrason. Symp., 435 (1990).

[17] A. Santoro, R.S. Roth, and M. Austin, Acta Crystallogr. B 38, 1094 (1982).

[18] S.C. Abrahams, W.C. Hamilton, and J.M. Reddy, J. Phys. Chem. of Solids 27, 1013 (1966).

[19] A. Bartasyte, V. Plausinaitiene, A. Abrutis, T. Murauskas, P. Boulet, S. Margueron, J. Gleize, S. Robert, V. Kubilius, and Z. Saltyte, Appl. Phys. Lett. 101, 122902 (2012).

[20] B.C. Grabmaier, W. Wersing, and W. Koestler, J. Cryst. Growth 110, 339 (1991).

[21] H. Bakhtaoui, Q. Micard, A. Mosset, V. Laude, K. Lengyel, G. Dravecz, P. Boulet, F. Sthal, G. Chevallier, A. Bartasyte, S. Margueron, J. Appl. Phys. 140 (11), 115106 (2026).